\documentclass[]{spie}  
\usepackage[]{graphicx}

\title{The Extreme Physics Explorer}

\author{Martin Elvis\supit{a}, 
\skiplinehalf
\supit{a}
Harvard-Smithsonian Center for Astrophysics,
60 Garden St., Cambridge MA 02138, USA}

\authorinfo{Further author information: (Send correspondence to
M.E.)\\
E-mail: elvis@cfa.harvard.edu, Telephone: +1 617 495 7442}

\begin{document} 
\maketitle 

\begin{abstract}

Some tests of fundamental physics - the equation of state at
supra-nuclear densities, the metric in strong gravity, the effect of
magnetic fields above the quantum critical value - can only be
measured using compact astrophysical objects: neutron stars and black
holes.
The Extreme Physics Explorer is a modest sized ($\sim$500~kg) mission
that would carry a high resolution ($R\sim$300) X-ray spectrometer and
a sensitive X-ray polarimeter, both with high time resolution
($\sim$5~$\mu$s) capability, at the focus of a large area
($\sim$5~sq.m), low resolution (HPD$\sim$1~arcmin) X-ray mirror.
This instrumentation would enable new classes of tests of fundamental
physics using neutron stars and black holes as cosmic laboratories.

\end{abstract}


\keywords{X-ray optics, spectroscopy, polarimetry, timing, neutron
  stars, black holes, magnetars, fundamental physics}

\section{INTRODUCTION}
\label{sect:intro}  

All scientific fields go through three phases: {\em discovery}, {\em
understanding}, and {\em tool}. In the {\em discovery} phase there is
widespread excitement over the new field and prizes are won as the
basic properties of the field are explored; in the {\em understanding}
phase interest narrows to those specialists willing to commit to the
intensive research that leads to physical understanding; at some point
though the remaining problems become so small that the field becomes
boring; not until the experts have achieved that understanding does the
field become of wide interest again, but this time as a {\em tool} to
unlock other fields of study.

X-ray astronomy is now over 40 years old and the first area of study
within X-ray astronomy - the neutron star and black hole X-ray
binaries - have long been in the {\em understanding} phase, and
interest in the field has shrunk to a relatively small expert
group. The progress made by this group however, largely thanks to the
{\em Rossi X-ray Timing Explorer} (Swank 1998), has been impressive,
so that many of the key astrophysical properties of these exotic
systems have been pinned down.
This progress has paved the way for the opening of the third phase:
the use of compact X-ray binaries as a {\em tool} to explore the {\em
extreme physics} for which neutron stars and black holes are natural
laboratories.

Here I propose the {\em Extreme Physics Explorer}, a moderate cost,
moderate risk, mission to allow precision measurements of the
fundamental physics that is only probed with these exotic objects.

\section{Extreme Physics}

It is well known that astrophysics provides many environments far more
extreme than can be created in laboratories on Earth. The early
universe is the best known case, and the discovery of Dark Energy
(Riess et al., 1998, Perlmutter et al., 1998) and the possibility of
exploring the nature of inflation via polarization signatures in the
Cosmic Microwave Background (Seljak \& Zaldarriaga 1997, Kamionkowski
et al., 1997) are exciting and vibrant fields of study.

The physics found in the extreme conditions of density found in
neutron stars, of strong gravity around black holes, and of quantum
critical magnetic fields in the magnetar class of neutron stars, while
studied, is not so dynamic a field. (The properties of each class of
object are described briefly below.) The reason for this lack of
dynamism is twofold: the emergence only recently of a good physical
understanding of much of the gas dynamics within these systems (the
astrophysics), which obscures the fundamental physics, and the lack of
precision tools with which to probe this physics. Most proposed
physics tests using these objects employ {\em timing} (Kouveliotou et
al., 1999, Heyl \& Hernquist 2003, DeDeo \& Psaltis 2004), but the
diagnostic potential of {\em high resolution spectroscopy}, though
less explored, is far greater. A combination of {\em time resolved
spectroscopy} would be most powerful of all. A fascinating set of
proposed physics tests use the {\em polarization} of the X-ray
radiation, e.g. as a test of the black hole metric (\S 2.2).

Both spectroscopy and polarization are less developed because both are
photon-hungry applications, making sufficiently sensitive observations
appear to be out of reach. The {\em Extreme Physics Explorer} makes these tests
practical.

\subsection{'Neutron' Stars}

Neutron stars are particularly promising objects with which to study
extreme physics as they:
\begin{enumerate}
\item probe regions of space-time curvature orders of magnitude more
  curved than e.g. the Hulse-Taylor pulsar (de Deo \& Psaltis, 2003);
\item are strong X-ray sources (10$^3$counts s$^{-1}$m$^{-2}$ for
  objects in our Galaxy), when in compact binary systems;
\item have a hard surface, enabling precision measurements;
\item rotate rapidly (P=1~msec - 1~minute), providing a reliable
  pulsed signal that enables time-dependent features to be resolved;
\item have a thin ($\sim$1~cm scale height) atmosphere, which imprints
  narrow atomic features in their spectra;
\end{enumerate}

All of these features combine to make them, in principle, excellent
'accelerators' with which to perform extreme physics experiments via
timing and spectroscopy.

\subsubsection{'Neutron' Star Equation of State}

Neutron stars are not likely to be made simply of neutrons at their
cores. Instead a variety of different particle compositions, all the
way down to the densest 'quark' stars have been proposed (Lattimer \&
Prakash 2001).  The way to distinguish these models, and so learn the
behavior of matter at extreme densities is via the equation of state
of neutron stars - i.e. their pressure-density relation. Although this
is a fifty year old question, this equation of state is at present
unknown.  The reason is that the relation depends on physics that
cannot be tested in the laboratory (Miller 2003). If the Mass-Radius
relation of 'neutron' stars could be measured, this would determine
their equation of state.  However, although orbital solutions for
X-ray binaries can measure the mass of the 'neutron' star, they do not
measure its radius.


The discovery by Cottam, Paerels \& Mendez (2002) using the Reflection
Grating Spectrometer (RGS, den Herder et al., 2001) on XMM-{\em
Newton} (Jansen et al., 2001) of a redshifted system of absorption
lines (Fe~XXVI and Fe~XXV 2-3, and OVII Ly$\alpha$) in the X-ray
spectrum of the binary EXO~0748-676 makes a strong case for a
gravitational origin of the redshift. With a value of 0.35$\pm$0.04
the redshift is consistent with neutron stars made of normal nuclear
matter, and excludes some more exotic states of matter such as strange
quark matter or kaon condenstates.

This is a breakthrough result in several ways: first it shows that
neutron star atmospheres, under some conditions, contain sharp
spectral signatures; second it shows that the redshift gives an M/R
value in the expected range, meaning that the origin is likely
gravitational. Cottam et al. thus opens up the whole field of
precision measurements of neutron stars.

However, the Cottam et al. result, unsurprisingly for a first
measurement, is limited in several ways:
\begin{enumerate}
\item The redshift has 10\% errors. This is not a limitation of the
  RGS, which has a spectral resolution $\lambda/\Delta\lambda$ =
  400. Hoogerwerf et al. (2004) have made a measurement with the {\em
  Chandra} LETGS, which has similar spectral resolution, accurate to
  $\pm$4~km~s$^{-1}$ (albeit for a relative velocity rather than an
  absolute one). Rather it is a limitation of counting statistics.
\item The measurement was integrated over 28 bursts from
  EXO~0748-676. Any changes in the emitting region from burst to
  burst, or around the 3.82~hr orbital period, could weaken and smear
  out the signal.
\item The observation was limited to the 3~s time resolution imposed
  by the readout of the CCD detectors of the RGS. All phase resolved
  information about the absorption lines is thus lost. 
\end{enumerate}

To measure M/R$\sim$0.3 to 1\% implies a spectral resolution,
$\lambda/\Delta\lambda$, $\sim$300, which is $\Delta$E=20~eV at 6~keV
and $\Delta$E=3~eV at 1~keV. The Cottam et al. result used lines close
to 1~keV, so 3~eV is a good fiducial $\Delta$E for any follow-up
spectroscopy.

Time resolved spectroscopy enables an entirely new constraint on the
neutron star radius. The Doppler shift of the absorption line changes
as it rotates around the neutron star, assuming that the ionized gas
in the atmosphere is localized rather than uniformly spread. For a
10~km radius and a 10~ms period this velocity would be of order
$\pm$6~km~s$^{-1}$. This would be a relative measurement and at the
level of precision attained by Hoogerwerf et al. would already yield a
3$\sigma$ detection. With excellent statistics and careful calibration
a precision amplitude could be measured. Then, combined with a spin
period for the neutron star known from light pulsations, the radius is
determined, modulo only the unknown $sin~i$ due to the inclination of
the neutron star rotation to our line of sight (but see \S 2.3).

\subsection{Black Holes}

The X-ray emission from black hole binary systems and active galaxies
originates close to the event horizon. Evidence for a strongly
gravitationally redshifted emission line due to the 6.4~keV neutral
Fe-K transition has been reported in both types of system (Tanaka et
al., 1994, Wilms et al., 2001, Miller et al., 2002, Tiengo et al.,
2005). In principle these lines can be used to probe the metric at the
strongest curvatures, next to the event horizon (Reynolds et al.,
2005).

These observations, though, are at the limit of count rate and
spectral resolution of current instruments, leading to tantalizing
results.  The variability properties of the broad redshifted component
of the Fe-K line are puzzling, but one explanation would put the
emission at just 2 Schwartzschild radii (Fabian 2005). Rapid
variability of narrow features within this broad line wing may be
signaling orbiting hot-spots with coherence times of a few orbits
(Turner et al., 2006). {\em Extreme Physics Explorer} spectra would
have the signal-to-noise, spectral resolution and bandwidth to turn
these edge-of-detectability results into tools.

Polarimetry offers another way to probe the warped metric near a black
hole. As different energy radiation is believed to originate from
different radii, the effects of GR warping of the light will be energy
dependent, and the size of the effect can determine the black hole
spin (Stark \& Connors, Connors \& Stark 1977, Connors, Piran \& Stark
1980 ApJ, 235, 224). Moreover the polarization angle of the radiation
determines the orientation of the accretion disk on the sky (Sunyaev
\& Titarchuk 1985). Since the black hole mass is known from binary
orbit analysis (or from reverberation mapping for AGNs), all of the
black hole parameters (except charge) are will then be determined.

The combination of spectroscopy and polarimetry over-determine the
black hole properties and so allow tests of GR in the strong gravity
regime. 

\subsection{Magnetars}

The 'magnetars' are a small subset of neutron star binaries containing
highly magnetized neutron stars. In these systems the magnetic field
is above the quantum critical limit $B_c$= $2\pi
m^2_ec^3/he$=4.4$\times$10$^{13}$~gauss, reaching much higher values
up to 10$^{15}$~g (Thompson \& Duncan 1995, 1996). At these field
strengths interesting new effects are predicted (Erber 1966, see
e.g. Miller 2001 for a recent review).
For example, the transmission of light through a super-critical field
is polarization dependent (e.g. Broderick \& Blandford 2003), and a
characteristic polarization signal is expected. This technique can
also determine the orientation of the neutron star rotation axis on
the sky (Meszaros et al., 1988), so removing one of last astrophysical
unknowns from the problem.

A spectroscopic example is that, at super-critical field strengths,
the cyclotron radius is smaller than the Bohr radius. As a result
there is a squeezing of the wave packet along the field lines. This
squeezing shifts the atomic energy levels, leading to e.g. hydrogen
Lyman-$\alpha$ being shifted up into the soft X-ray regime (Ho et al.,
2003).

\subsection{Astrophysics}

Although the {\em Extreme Physics Explorer} is designed as a physics
experiment, it will also be powerful as an observatory for a wide
range of astrophysics, and we list a few examples here:

\noindent{\em Spectroscopy:}
\begin{itemize}
\item 'occultation' experiments via rapid variability of absorbers in
  AGNs (Risaliti et al., 2006);
\item AGN wind location and characterization using non-equilibrium
  modelling of 'Warm Absorbers' (Krongold et al., 2006);
\item spectral Doppler mapping of the nearest several supernova
remnants, galaxies and clusters of galaxies;
\item thermonuclear burning in X-ray bursts;
\item element abundances in star formation regions;
\item stellar coronal dynamics;
\item charge exchange spectra of comets;
\end{itemize}

\noindent{\em Polarization:}
\begin{itemize}
\item reflection spectra of hidden AGNs;
\item emission from white dwarf binaries (CVs);
\item accretion disk atmosphere scattering;
\item diagnostics of jets in quasars and microquasars;
\item scattered light in clusters of galaxies (Churazov 2003).
\end{itemize}

\section{The {\em Extreme Physics Explorer}} \label{sect:sections}

To overcome the limitations of current, and currently planned,
observations requires instrumentation with much larger area, much
improved time resolution and at least comparable spectral resolution
to those now flying. An advance of this magnitude is possible using
microcalorimeters as high time resolution spectrometers, combined with
existing extremely light-weight X-ray optics to give large collecting
areas. The area needed is of order a few square meters: each
observation needs 10$^8$ counts, while most X-ray binaries count at
10$^3$ct~s$^{-1}$m$^{-2}$; a measurement needs to take place in a time
small compared to an orbital period, or to changes in the
astrophysical gas flow, i.e. hours or less. Such large areas preclude
the use of dispersive spectroscopy (gratings), as these need good
optics (HPD $<$ 10~arcsec).

The essential realization behind the {\em Extreme Physics Explorer} is
to note that, in this special application, 'bad imaging is good'; that
is a diffuse point spread function (PSF) can be used to spread the
signal over an 1000 pixel microcalorimeter array, enabling the array
as a whole to count at 1~MHz, as needed.

\subsection{Microcalorimeters}

Microcalorimeters are not usually thought of as good timing devices.
Yet the rise time of the pulse created by an individual photon in a
microcalorimeter is only $\sim$50~$\mu$s, so that the pulse peak can
be timed to $\sim$5~$\mu$s.  Since microcalorimeters are now
approaching an energy resolution, $\Delta$E$\sim$2~eV, and have
essentially unit quantum efficiency down to at least 0.5~keV, they
would appear to be ideal detectors for the study of physics using
compact objects.

Their limitation is that the decay time of the pulse is much longer,
$\sim$300~$\mu$sec (e.g. Ohno et al., 2002). Hence pulses arriving at
rates greater than $\sim$1~kHz will pile-up on one another and the
energy resolution of the detector will be degraded. Con-X will suffer
from this limitation. 1000~ct~s$^{-1}$ is too low a rate to allow
interesting timing analysis.

Fortunately, this is a {\em per pixel} count rate limit. A uniformly
illuminated 32$\times$32 microcalorimeter array could count at 1~MHz,
a rate adequate to gather the needed signal in the requisite time.
However, microcalorimeter arrays are small, $\sim$1~cm$^{2}$, and need
to be cooled to $\sim$70~mK, so that scaling up to square meter areas
is unrealistic. Clearly the microcalorimeter array needs to be fed by
a large, though not exquisite, X-ray optic.

As shown in section~3.3, the {\em Extreme Physics Explorer} will have
a plate scale of 1.5~cm/arcmin, so that a pixel size of
$\sim$500~$\mu$m (2~arcsec) is needed to get the required per pixel
count rate. 500~$\mu$m pixels are double the dimension of those
planned for Con-X (Kilbourne et al., 2005) and the same as those for
DIOS (Ohashi et al., 2006). So this is a good match of requirements to
technology.

To achieve $\Delta$E = 3~eV needs a small heat capacity in the
pixel. Compared to Con-X, a thinner converter could be used. This
would lower the heat capacity, as needed, but would also lower the
quantum efficiency of the array at high energies. A solution to this
involving a 'beam splitter' is described below (\S 3.3).

The minimum required array size is 32$\times$32, but this allows for
no jitter in the image location on the detector, and so would lead to
unreasonably tight tolerances on the aspect solution and the
mirror-detector alignment. A 42$\times$42 array would allow for
5~arcsec rms image jitter which is a more realistic minimum
requirement.  A 64$\times$64 array would leave much more leeway for
the rest of the system, but is just beyond current fabrication
capabilities. The trade between array size and optical bench
tolerances is non-obvious and will need study.

\subsection{Polarimeters}

The polarization of cosmic X-ray sources is a poorly developed field.
Only the Crab Nebula has a detected polarization signal (Weiskopf et
al., 1978). The underlying cause for this lack of development is the
photon-hungry nature of polarimetry: four Stokes parameters need to be
determined, and to detect a 1\% polarization signal at 10~$\sigma$
requires 10$^6$ photons. This calculation assumes 100\% modulation of
the signal, which is never achieved in real devices. The 'minimum
detectable fractional polarization' (at 99\% confidence), $M=
\frac{4.29}{\mu S}\surd \frac{(S+B)}{t}$, where $S$ is the source
count rate, $B$ the background count rate, $t$ the observing time, and
$\mu$ the modulation factor. {\em Extreme Physics Explorer} provides
large $S$ and small $B$. The crucial remaining parameter is $\mu$,
which must be close to unity, but has not been the case for
polarimeters in the past.

The potential for X-ray polarimetry is, however, large. This was
demonstrated by the 2004 'X-ray Polarimetry Workshop' at the Kavli
Institute for Particle Astrophysics and Cosmology\footnotemark. Many
applications of X-ray polarimetry were discussed, and a number of
novel techniques for detecting polarization were presented at this
meeting.

\footnotetext{URL: http://www-conf.slac.stanford.edu/xray\_polar/Talks.htm}

One of the most promising polarimeter developments for the standard
energy range of grazing incidence optics (0.1-10~keV), is that of the
'Micro Pattern Gas Polarimeter' (Costa et al., 2001).  This device has
several characteristics that are well suited to the Extreme Physics
Explorer. In this device polarization information is derived from the
tracks of the photelectron, which are imaged using a finely subdivided
gas detector ('PIXI', Costa et al., 2006). The tracks are long enough
that a 50\% modulation factor has been achieved at 5.4~keV. The third
generation device now being tested is 1.5~cm diameter, well matched to
the PSF of the {\em Extreme Physics Explorer}.  The individual
triggering of the 10$^6$ pixels in this device enables the 3 -
10~$\mu$s time resolution appropriate for {\em Extreme Physics
Explorer}. The Micro Pattern Gas Polarimeter has the moderate energy
resolution typical of other gas counters, and operates at around room
temperature. Being a gas counter the instrument could not operate at
the cryogenic temperatures of a microcalorimeter, as the gas would
liquefy. Use of this device would thus require dividing the focal
plane into cold and warm sections. A compact cryostat (e.g. a 50~cm on
a side cube, as on DIOS, Ohashi et al., 2006), and the inherently wide
field opticcs of conical optics (\S 3.3), would allow the two
instruments to share the focal plane. Offset pointing would be used to
select which one recieves the target's photons.

\subsection{Microchannel Plate Optics}
\label{mcp}

Microchannel plate optics (Beijersbergen et al., 1999) have the right
properties to create the low resolution mirror that feeds the
instruments on the {\em Extreme Physics Explorer} \footnotemark:
\begin{enumerate}
\item weight 3.7~kg~m$^{-2}$: c.f. 250~kg~m$^{-2}$ for the Con-X optics;
\item sub-arcminute PSF: demonstrated by Bavdaz et al. (2002);
\item high aperture utilization, with good reflectivity up to high
energies;
\item plate-like thin, rigid structures: can fold and deploy readily,
  like solar panels;
\item large field of view: inherent in the conical approximation to
  Wolter~I optics (Petre \& Serlemitsos 1985).
\end{enumerate}

\footnotetext{Note that these are {\em not} the same technology as the
  micropore silicon optics being developed for XEUS (Collon et al.,
  2006) which are heavier, but have much higher resolution.}

These optics have been developed extensively by the University of
Leicester and ESA's ESTEC center, for use in the LOBSTER all sky
monitor program (Fraser, G. \& Bannister, N., 2002), based on an
original concept by Roger Angel (1979).
An arcminute PSF spreads out the signal over many pixels, as needed
for the {\em Extreme Physics Explorer}. It may be possible to optimize the
PSF further toward the ideal of uniform illumination, by introducing
small deliberate alignment offsets individual small plates in the
mirror. The tolerances required to produce a 'top hat' beam profile
need investigation.

While an arcminute-scale PSF is by no means high resolution, and in
fact performs the valuable function of spreading out the signal on the
microcalorimeter array, it is not so bad, being similar to that of
ASCA (Tanaka et al., 1994). The {\em Extreme Physics Explorer} would
resolve the 'nest of black holes' toward the Galactic Center (though
not the central black hole X-ray source Sgr~A*), and would be capable
of reaching the brightest few dozen Active Galaxies (AGNs) and quasars
without being excessively background limited. (The use of active
anti-coincidence shields around the microcalorimeter on {\em Suzaku}
led to lower backgrounds than in the ASCA CCDs, Kilbourne et al.,
2006.) A few additional pixels unexposed to the incoming X-ray beam
could serve as a recorder of the remaining background. Background is
unimportant for Galactic X-ray binary observations.

To reach a collecting area of several square meters requires a
$\sim$5~m diameter mirror. This might be achieved with a set of seven
1.7~m diameter mirror panels that deploy in orbit. To preserve the
small, grazing incidence, angles needed for X-ray reflection, an long
focal length, $\sim$40~m, is required. At 40~m a 1~arcmin spot has a
size of 1.5~cm. This is a good match to microcalorimeter arrays.
The 40~m long focal length is much less of a difficulty than might be
imagined (see next section), and is eased by the cm-sized spot created
by the microchannel plate optics, as this leads to relatively loose
tolerances for maintaining the relative mirror-detector location.

Microchannel plate optics have good high energy reflectivity, so it
would be wasteful not to detect these photons efficiently, not least
because the 6.4~keV Fe-K line is an important diagnostic around black
holes. Yet to attain $\Delta$E = 3~eV at 1~keV, the microcalorimeter
array may have to compromise high energy quantum efficiency. An ideal
arrangement would be to have a beam splitter to divide the incoming
X-ray beam by energy at $\sim$2~keV. The two beams could then be
directed to separate microcalorimeter arrays optimized for the two
bands. 

I propose here a novel way of creating an X-ray beam splitter: by
having the outer parts of the second reflector - which have larger
graze angles and reflect lower energy X-rays - canted by a few
($\sim$5) arcminutes from optical axis of the inner parts of the
mirror. (Normally one would talk of the outer and inner shells, but
microchannel plate optics do not have simple shells.) This small tilt
is only 10\% of the graze angle, so will have minimal effect on the
reflectivity, while the PSF is dominated by alignment effects and will
not be affected by this small off-axis deviation. A side benefit of
separated high and low energy response is a doubling of the maximum
mission count rate. If the tilt is chosen so that the two PSFs overlap
slightly then a small fraction of the low energy beam will fall on the
high energy detector, and vice versa. This will enable good
cross-calibration of the two arrays, as will the overlap in their
effective area vs. energy curves.

\subsection{Spacecraft}

Clearly a 40~meter focal length requires an extending optical
bench. (Launcher farings do not exceed lengths of a few meters.) The
alternative of a pair of separate, independently free-flying, mirror
and detector spacecraft, as planned for {\em Simbol-X} (Ferrando et
al., 2005) adds major complications, especially in a low earth orbit
where gravity gradients are strong.

A minimum factor of $\sim$10 extension is needed for the optical
bench. Fortunately, a light-weight deployable optical bench exists and
has been flight-tested on numerous missions: UARS, GGC WIND, GGS
POLAR, Cassini, Lunar Prospector, IMAGE (Able Engineering: URL
http://www.aec-able.com/Booms/ablebooms.html,
http://www.aec-able.com/ableflightherita.html). Most of these were
relatively short booms, but the longest reached $\sim$20~meters. An
extension to 40~meters should not be problematic.

Thanks primarily to the lightweight optics and optical bench, the
total mass of the {\em Extreme Physics Explorer} is modest,
$\sim$500~kg.  Table~1 gives a preliminary breakdown of the mass
budget. The calorimeter/Cryostat mass may be reduced somewhat if a
DIOS-like design (100~kg, Ohashi et al., 2006) is adopted. A modest
additional mass for the polarimeter instrument needs to be calculated
and included.

\begin{table}[t]
\caption{Mass Budget}
\begin{center}
\begin{tabular}{|l|r|}
\hline
Item&Mass (kg)\\
\hline
10~m$^2$ microchannel plate mirror& 40\\
Mirror support assembly& 40\\
Optical bench, extending to 40~m& 40\\
Optical bench canister& 50\\
Calorimeter \& Cryostat$^*$& 125\\
Polarimeter& 30 \\
'All Sky' monitor& 50\\
Spacecraft & 200\\
20\% reserve & 105\\
\hline
{\bf Total}&{\bf 690}\\
\hline
\end{tabular}

$^*$ Based on ASCA design.
\end{center}
\end{table}

The operations of the {\em Extreme Physics Explorer} are expected to
consist primarily of $\sim$day-long pointings at bright (10$^3$
ct~s$^{-1}$~m$^{-2}$) Galactic X-ray binaries. A long ($>$5 year)
mission is feasible if a cryogen-free cryostat can be employed
(e.g. the 5-stage system proposed for DIOS, Ohashi et al., 2006).
Rapid slews to ToOs are implausible with a 40~m long spacecraft.  A
small 'all sky monitor' (e.g. ASPEX, Feroci et al., 2006), to alert
mission controllers to transient X-ray sources, would be a useful
addition to the mission, as these sources remain bright for days to
weeks, and track through a range of accretion rates, some of which are
more optimal than others for physics experiments.

The low mass of the {\em Extreme Physics Explorer} leaves adequate
launch capability to reach a high orbit.  A high orbit would enable
continuous viewing of these targets which: (1) simplifies Fourier
analysis of their properties, (2) ensures full coverage of orbital
phases, (3) doubles the observing efficiency, compared with LEO, (4)
reduces gravity gradient torques, simplifying the attitude control
system; and (5) reduces the heat load on the cryostat.

This observing profile implies continuous event rates of
10$^4$~ct~s$^{-1}$. Each microcalorimeter event will need 10 energy
bits (2~eV in 2~keV), 20 timing bits (5~$\mu$s in 5~s), and 10 pixel
ID bits, which sum to: 4$\times$10$^5$~baud. Including aspect and
housekeeping a continuous data rate of 0.5~MB is needed. Ground
contact will need to be extensive. Perhaps even a GEO orbit should be
considered because of the continuous coverage it affords, despite the
high backgrounds encountered there.

\section{Main Challenges}

For convenience we summarize the main challenges to realizing the
{\em Extreme Physics Explorer}:
\begin{itemize}
\item Microcalorimeter arrays with $<$3~eV resolution and at least
  42$\times$42 format (64$\times$64 preferred), and the multiplex
  readout electronics to support them;
\item Mass manufacture of microchannel plate optics, and their alignment;
\item Deployment of MCP optics and optical bench, mirror thermal control;
\item Pointing and stability of optical bench;
\item Size of cryostat (to allow polarimeter close to optical axis);
\item High data rate (background in GEO?);
\item Refinement of science case.
\end{itemize}

This last item, the refinement of the science case, is pressing as it
may alter the mission architecture. The full range of potential
physics tests has surely not been exhausted in the literature.  The
science case needs to be carried through to full simulations of
observations of specific X-ray binaries. Targeted workshops to
develop the science case for the {\em Extreme Physics Explorer} would be
valuable.

\section{Conclusions}

The {\em Extreme Physics Explorer} is a modest sized ($\sim$500~kg) mission
that would carry an X-ray spectrometer and an X-ray polarimeter, both
with high time resolution ($\sim$5~$\mu$s), at the focus of
a large area ($\sim$5~sq.m), low resolution (HPD$\sim$1~arcmin) X-ray
mirror.

This instrumentation would enable a new class of tests of fundamental
physics using neutron stars and black holes as cosmic laboratories.

\acknowledgments
I thank George Fraser, Gareth Price, Jay Bookbinder, Rich Kelley,
Caroline Kilbourne, Enrico Costa, Takaya Ohashi, Simon De Deo,
Nancy Brickhouse and  Frits Paerels for valuable discussions.




\end{document}